\documentclass[preprint,aps,pre,showpacs,eqsecnum,nobibnotes,amssymb,amsmath,floatfix]{revtex4}
\usepackage{graphicx}
\usepackage{fontenc}

\renewcommand{\imath}{{\rm i}}

\begin{document}
%%%\texttt{graphics}

%****************************************************************
\title{Noise sensitivity of sub- and supercritically bifurcating patterns 
with group velocities close to the convective-absolute instability\\}

\author{A.~Szprynger}
\affiliation{Institute of Low Temperature and Structure Research,
Polish Academy of Sciences, POB 937, 51-112 Wroc\l aw, Poland}
\author{M.~L\"ucke}
\affiliation{Theoretische Physik, Universit\"at des Saarlandes,
Postfach 151150, D-66041 Saarbr\"ucken, Germany}
\date{\today}
%****************************************************************

\begin{abstract}
The influence of small additive noise on structure formation
near a forwards and near an inverted bifurcation as described by a cubic and
quintic Ginzburg Landau amplitude equation, respectively, is studied numerically
for group velocities in the vicinity of the convective-absolute instability 
where the deterministic front dynamics would empty the system.
\end{abstract}

%  47.20.Ky :  Nonlinearity (including bifurcation theory)
%  47.54.+r :  Pattern selection; pattern formation
%  43.50.+y :  Noise: its effects and control
%  05.40.-a :  Fluctuation phenomena, random processes, noise, and Brownian motion
\maketitle
%\tableofcontents

%%\narrowtext
%\clearpage
%%%%%%%%%%%%%%%%%%%%%%%%%%%  Sec. I %%%%%%%%%%%%%%%%%%%%%%%%%%%%%%%%
\section{Introduction}
%%%%%%%%%%%%%%%%%%%%%%%%%%%%%%%%%%%%%%%%%%%%%%%%%%%%%%%%%%%%%%%%%%

The formation of macroscopic structures \cite{CH93} in systems that
are driven out of thermal equilibrium by an externally imposed generalized
stress are usually investigated by deterministic field
equations. However, under specific circumstances the influence of external
deterministic or stochastic perturbations
and of internal thermal noise on the pattern formation process should be taken into
account to achieve a more realistic and quantitative
description of experiments. One prominent example are the so-called noise
sustained structures
\cite{D85,BAC,Steinberg,MLK92,SR92,LR93,SBH94,Deissler94,NWS96,LS97,T97,SCMW97,CWS99}
in the convectively unstable parameter regime
\cite{BersBriggs,Huerre} in, e.g., the Taylor-Couette
\cite{BAC,Steinberg,LR93,SBH94,Deissler94},
the Rayleigh-B\'enard \cite{MLK92,SR92,T97} system, or nonlinear optics 
\cite{SCMW97}. Further examples are certain open-flow instabilities , e.g., in 
wakes and jets that are reviewed in \cite{Huerre}.

The noise sustained structures
\cite{D85,BAC,Steinberg,MLK92,SR92,LR93,SBH94,Deissler94,NWS96,LS97,T97,SCMW97,CWS99}
arise when an externally imposed through-flow or an internally generated
group velocity $v$ is large enough to "blow"
the pattern out of the system according to the
deterministic field equations. In this driving regime one observes in
experiments \cite{BAC,Steinberg,SR92,T97,SCMW97} structures that are sustained by
sources that generate perturbations in the band of modes that are amplified
according to the supercritical deterministic growth dynamics in downstream
direction sufficiently far away from the inlet.

The criterion \cite{BersBriggs,Huerre} at which $v$
the pattern is blown out of the system under deterministic laws which gave
the threshold for the appearance of the noise sustained, supercritically
bifurcating patterns in the above described experiments is a linear one. It was
nonlinearly extended by Chomaz \cite{Chomaz92} to the question of the
propagation direction of nonlinear
deterministic fronts in infinite systems that connect the unstructured state
to the finite-amplitude structured one.

Here we study and compare the noise sensitivity of pattern forming systems in
which the above described fronts are linear or nonlinear ones. To that end we
investigate the cubic Ginzburg-Landau amplitude equation (GLE) for a supercritical
forwards bifurcation and the quintic GLE for a subcritical inverted bifurcation,
respectively, in one spatial dimension.

We solve the GLE with additive stochastic forcing numerically. Our systems
are finite but sufficiently long to allow the establishment of a statistically
stationary large-amplitude bulk part -- provided the latter is possible with the
boundary condition of vanishing amplitude at the ends. We focus our attention to
parameters in the vicinity of the convective-absolute threshold at which the 
fronts of the
deterministic GLE cease to propagate. And we investigate in particular the
statistical dynamics of phase and amplitude fluctuations in the front region.

%\clearpage
%%%%%%%%%%%%%%%%%%%%%%%%%%%  Sec. II %%%%%%%%%%%%%%%%%%%%%%%%%%%%%%%%
\section{SYSTEM}
%%%%%%%%%%%%%%%%%%%%%%%%%%%%%%%%%%%%%%%%%%%%%%%%%%%%%%%%%%%%%%%%%%
We consider the stochastic, 1D Ginzburg-Landau equation
\begin{equation}
(\partial_{t}+v\partial_{x}) A\;=\;(\mu + \partial_{x}^2
+ g_3 |A|^2  + g_5 |A|^4 ) A + \sigma \eta
\label{CGLE}
\end{equation}
for the complex amplitude
\begin{equation}
A\;=\;\Re A + \imath\, \Im A\;=\;R e^{\imath\,\Phi}
\label{A_ReIm}\end{equation}
depending on $x,t$. Here $\Re (\Im)$ denotes the real (imaginary) part
and $R=|A|$ is the modulus and $\Phi$ the phase of $A$.
The coefficients in (\ref{CGLE}) are taken as real for simplicity. We checked
however that taking into account the (small) imaginary parts, that appear
e.g.~in the case of transverse Rayleigh-B\'enard convection rolls propagating
downstream in a small externally imposed lateral through-flow \cite{MLK92} or in
the case of downstream propagating Taylor vortices \cite{BAC,RLM93} does not
change the major findings presented in this paper significantly.
We consider the group- or mean flow velocity $v\ge 0$ in positive $x$-direction
and the linear growth rate $\mu$ of $A$ as control parameters.

We investigate
two fixed combinations of the nonlinear coefficients ($g_3,g_5$) that we refer
to in this paper as follows
\begin{subequations}
\begin{eqnarray}
g_3\;=\;-1\;, \; g_5\; = \;0\; &:&  \quad \mbox{cubic GLE} \\
g_3\;=\; 1\;, \; g_5\; = \;-1\;&:&  \quad \mbox{quintic GLE} \, .
\end{eqnarray}
\end{subequations}
The quantity $\sigma$ in (\ref{CGLE}) measures the real strength of the complex
stochastic force
\begin{equation}
\eta(x,t)\;=\;\Re\eta(x,t) + \imath\, \Im\eta(x,t)
\end{equation}
with statistically independent real and imaginary parts $\Re\eta$ and $\Im\eta$,
respectively. Both are Gaussian distributed with zero mean and
$\delta$-correlated such that
\begin{equation}
<\eta(x_1,t_1)[\eta(x_2,t_2)]^*>\;=\;2\delta(x_1-x_2)\delta(t_1-t_2) \;.
\label{f_corr}
\end{equation}
%%%%%%%%%%%%%%%%%%%%%%%%%%%%%% Sec IIA  %%%%%%%%%%%%%%%%%%%%%%%%
\subsection{Unforced homogeneous solution}
%%%%%%%%%%%%%%%%%%%%%%%%%%%%%%%%%%%%%%%%%%%%%%%%%%%%%%%%%%%
We are interested in the effect of small additive noise on
the spatio-temporal structure formation in large but finite or semiinfinite
systems. Nevertheless it is useful to briefly recall first the properties of the
most simple solutions of the unforced GLE in an infinite system. This shows what
one might expect to see in the bulk of a very large system far away from the
boundaries --- ignoring for the moment questions related to boundary induced
pattern selection processes.

The GLE (\ref{CGLE}) shows for $\sigma$=0 a continuous family of traveling wave
(TW) solutions
\begin{equation}
\label{TWsolution}
A(x,t)= R e^{iq(x-vt)}
\end{equation}
with constant wave number $q$, frequency $\Omega=-qv$, and modulus $R$ given by
\begin{equation}
\mu -q^2 + g_3 R^2 + g_5 R^4 = 0 \, .
\end{equation}
This TW solution family bifurcates at the marginal stability curve, $\mu = q^2$,
of the $A$=0 solution out of the latter while the former becomes unstable there.
The critical values are $\mu_c=q_c=\Omega_c$=0. The bifurcation is nonhysteretic
and forwards in the cubic case
\begin{equation}
\label{Rcubic}
R^2=\mu -q^2
\end{equation}
and hysteretic, backwards in the quintic case
\begin{equation}
\label{Rquintic}
R^2=\frac{1}{2} \pm \sqrt{\mu -q^2 + \frac{1}{4}} \, .
\end{equation}
Here the lower sign refers to the lower unstable TW solution branch that exists
for $-\frac{1}{4} \leq \mu -q^2 \leq 0$. The upper TW solution branch identified
by the + sign in Eq.~(\ref{Rquintic}) exists beyond the saddle-node bifurcation
value $\mu = q^2 -\frac{1}{4}$. These TW solutions are stable for wave numbers
outside the Eckhaus unstable band \cite{BD92}.

%%%%%%%%%%%%%%%%%%%%%%%%%%%  Sec. IIB %%%%%%%%%%%%%%%%%%%%%%%%%%%%%%%%
\subsection{Convective-absolute instability}
\begin{figure}
\includegraphics[clip=true,width=12cm,angle=0]{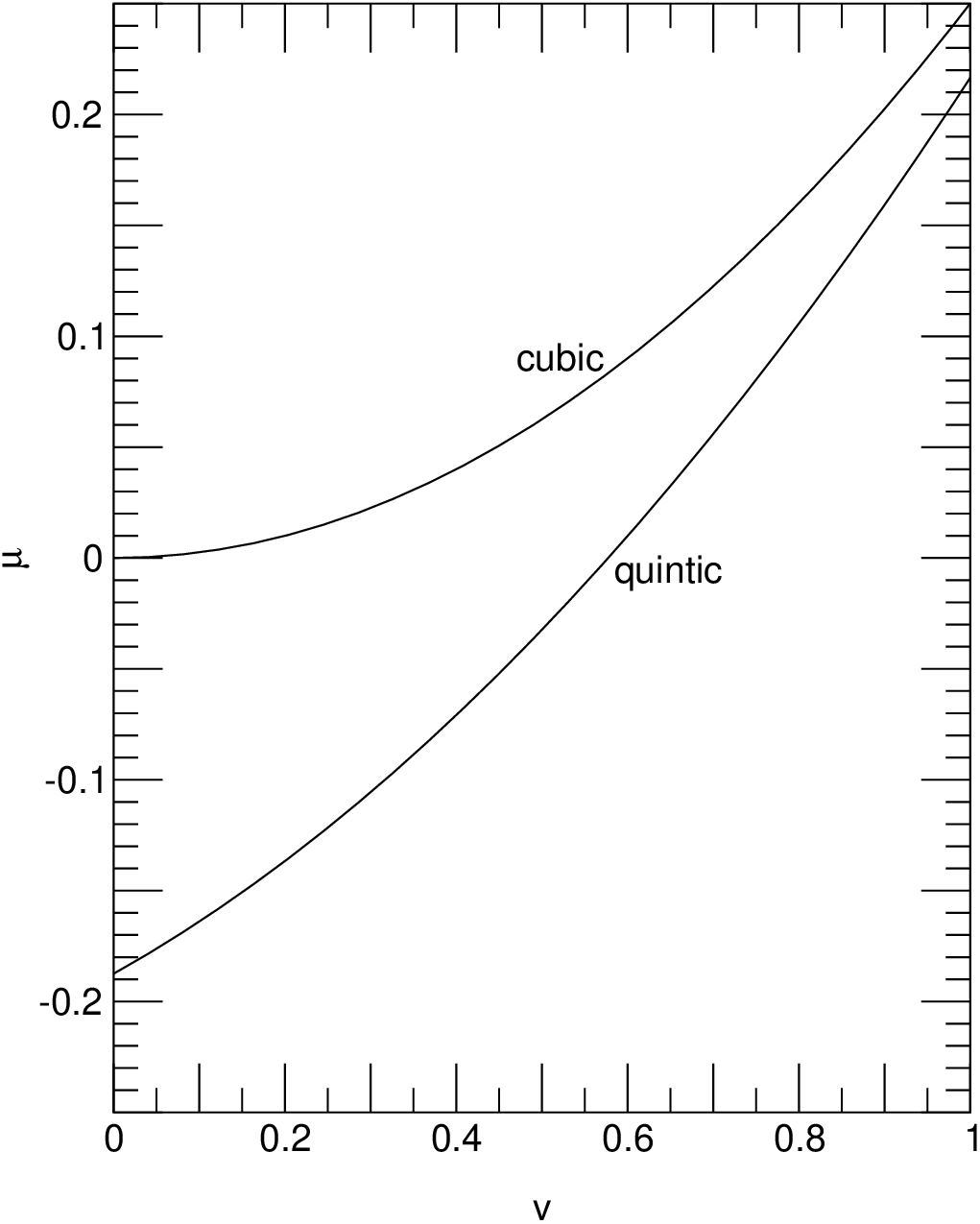}
\caption{\label{figure:Maxwell}
Convective-absolute instability boundaries (\ref{mu_conv}) for the unforced 
cubic and quintic
GLE, respectively. For parameters below the respective curve front propagation
is such that in the absence of noise the $A$=0 state invades the $A\neq$0 state.
In the absolutely unstable parameter regime above the respective curve the
$A$=0 state recedes and the $A\neq$0 state expands (as long as the front is not
hindered by a boundary).}
\end{figure}

%%%%%%%%%%%%%%%%%%%%%%%%%%%%%%%%%%%%%%%%%%%%%%%%%%%%%%%%%%%%%%%%%%
The noise susceptibility of the pattern formation process described by the GLE
(\ref{CGLE}) changes significantly \cite{D85,Huerre} when crossing the parameter
combination of
$\mu, v$ shown in Fig.~\ref{figure:Maxwell} for the so called
convective-absolute instability \cite{BersBriggs}. This combination
\begin{eqnarray}
\mu_{c-a}=
\left\{ \begin{array}{ll} \frac{1}{4} v^2  &\mbox{cubic GLE} \\
\frac{3}{16}\left(v^2 + \frac{2}{\sqrt{3}} v - 1 \right)\quad & \mbox{quintic GLE}
\end{array}\right.
\label{mu_conv}
\end{eqnarray}
is marked by the front solution of the deterministic GLE  with
$\sigma=0$ undergoing a reversal of the front propagating direction in an
infinite system. Consider a front that connects the basic state $A=0$ being
realized at $x\to-\infty$ to a homogeneous solution with $A\ne 0$ at
$x\to\infty$. For parameter values below (above) the respective curves in
Fig.~\ref{figure:Maxwell} this front moves to the right (left). Thus the basic
state $A=0$ (the homogeneous solution $A\ne 0$) expands to the right (left).
The region below (above) the respective curves in Fig.~\ref{figure:Maxwell}
where the basic state $A=0$ (the homogeneous state $A\ne 0$) invades the whole
system is called the convectively (absolutely) unstable region of the $A=0$
solution \cite{D85,Huerre}. Thus, the boundary (\ref{mu_conv}) is also called
the convective-absolute instability boundary.

For the cubic GLE the boundary $\mu_{c-a}=v^2/4$ results from a linear analysis
\cite{D85}.
For the backwards bifurcating solution arising in the quintic GLE the respective
front that reverts its propagation direction is a nonlinear one \cite{SH92}. Note
that in the latter case the convective-absolute instability boundary 
\cite{Chomaz92}
connects for $v\to 0$ to the so-called Maxwell point $\mu_{c-a}=\mu_M=-3/16$:
For this value the minima of the potential $V(A)=-\frac{\mu}{2}A^2 -
\frac{1}{4}A^4 + \frac{1}{6}A^6$ have equal height $V=0$.

\begin{figure}
\includegraphics[clip=true,width=9cm,angle=0]{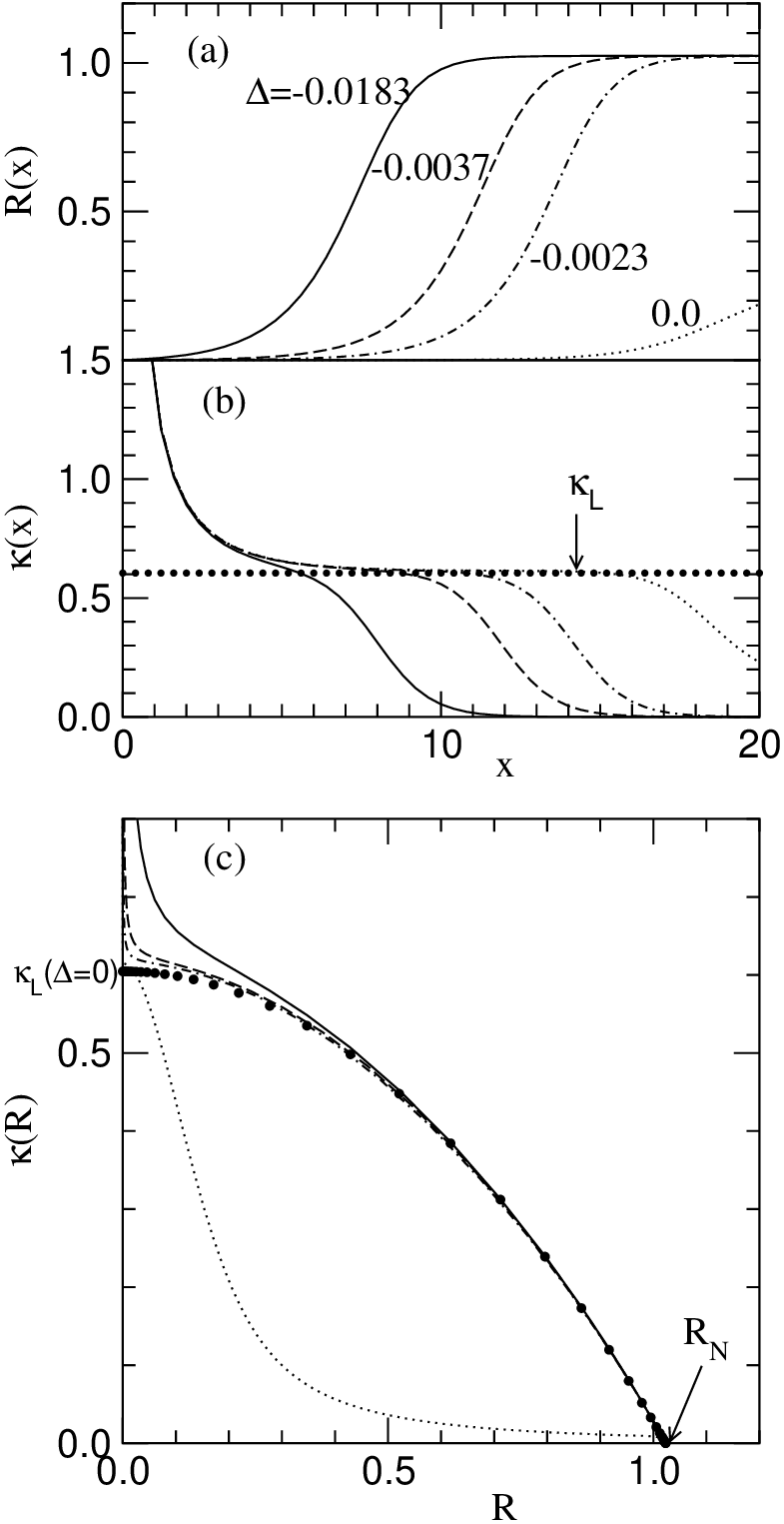}
\caption{\label{figure:R-kappa}
Deformation of the front solution $R(x)$ (a) of the
deterministic quintic GLE by the boundary condition $A(x=0,t)=0$ in the
absolutely unstable
regime for $\mu=0.05$ and $v=(1+\Delta)v_{c-a}$ as indicated.
The spatial growth rate $\kappa(x)=\partial_x ln{R(x)}$ (b) deviates from a
freely propagating front with wave number $q=0$ that would show \cite{SH92}
$\kappa_L=\frac{v}{2}+\sqrt{\frac{v^2}{4}-\mu}$ [thick dots in (b)] in the
small-amplitude "linear" part of the front.
(c) shows $\kappa$ versus $R$ in comparison with the prediction \cite{SH92}
$\kappa=\frac{1}{\sqrt{3}}\left(R_N^2 - R^2\right)$ for a
stationary front in an infinite system at $\Delta=0,\,\mu=0.05$ for which
$R_N^2$=1.048.
Thin dotted curves in (a)-(c) refer to a numerically obtained solution for
$\Delta=0$ at time $5\cdot 10^4$ which is not yet stationary. Here the profile is
still
moving to the right and in the absence of numerical "noise" we would expect this
transient to approach the $R\equiv 0$ basic state [c.f.~also Fig.~(c)].}
\end{figure}

The boundary condition $A(x=0,t)=0$ that we apply in our simulations stops
any front
propagating to the left and it changes, i.e., it deforms the front profile when
the front is sufficiently close to the boundary at $x=0$. This can be seen in
Fig.~\ref{figure:R-kappa} for the example of the deterministic quintic GLE. There
the lines show the modulus profile $R$ and the spatial growth rate $\kappa=R'/R$
versus $x$ together with $\kappa$ versus $R$ obtained numerically for several
parameter values above the convective-absolute instability boundary. To 
facilitate comparison of different cases we
introduce the reduced horizontal "distance"
\begin{equation}
\Delta\;=\;\frac{v}{v_{c-a}} - 1
\end{equation}
from the boundaries shown in Fig.~\ref{figure:Maxwell}. Here
\begin{eqnarray}
v_{c-a}(\mu)\;=\;
\left\{ \begin{array}{ll} 2 \sqrt{\mu} \quad &\mbox{cubic GLE} \\
\sqrt{\frac{4}{3} \left(1 + 4\mu\right)} - \sqrt{\frac{1}{3}}
\quad & \mbox{quintic GLE}
\end{array}\right. \label{v_c-a}
\end{eqnarray}
denotes the convective-absolute instability boundary (\ref{mu_conv}).

The results that we present here were obtained for $\mu>0$, i.e., in a situation
where the basic state $A=0$ is unstable. For the backwards bifurcation in the
quintic GLE with negative growth rates $ -\frac{3}{16} < \mu < 0$ for which the
above cited potential has a minimum at $A=0$ the situation is more complicated
\cite{CWS99}: Not only does the establishment of the final front connecting the
inlet condition $A=0$ with a statistically stationary saturated bulk with
$|A|={\cal O}(1)$ depend sensitively on the initial condition [say, $A(x,t=0)=0$
versus $|A|={\cal O}(1)$] in the absolutely unstable regime, $\Delta < 0$. But
more importantly, in the convectively unstable regime, $\Delta > 0$, we found that
small
noise does not seem to be able to generate with the boundary condition $A(x=0)=0$
a noise sustained finite-amplitude structure with $\langle |A|^2\rangle$ of order
one when $\mu<0$: The deterministic front dynamics drives the large-amplitude part downstream
and eventually any finite system is filled only with small-amplitude fluctuations
of $A$ around the stable fixed point $A=0$ of the unforced system.

%%%%%%%%%%%%%%%%%%%%%%%%%%%%%% Sec IIC  %%%%%%%%%%%%%%%%%%%%%%%%
\subsection{Noise strength}
\label{SEC:noise_strength}
%%%%%%%%%%%%%%%%%%%%%%%%%%%%%%%%%%%%%%%%%%%%%%%%%%%%%%%%%%%%%%%%%%

For the quintic GLE we choose the noise strength $\sigma=10^{-3}$. The noise
intensity $\sigma^2$ should be compared with the minimum of the potential
\begin{equation}\label{V_3-5}
V(A)=-\frac{\mu}{2}A^2 - g_3 \frac{1}{4}A^4 - g_5 \frac{1}{6}A^6 \, . 
\end{equation}
For our quintic case ($g_3=1, g_5=-1$) the minimum at 
$A^2=R_N^2=\frac{1}{2} + \sqrt{\frac{1}{4} + \mu}$ is 
$V(R_N)=-\frac{1}{24}[1 + 6\mu + (1 + 4 \mu)^{3/2}]$. Thus, the noise 
"temperature" $\sigma^2$ measured in units of $V(R_N)$ is $\sigma^2/|V(R_N)|=
9.2 \, 10^{-6}$
for the control parameter $\mu=0.05$ that we have used in most of our
calculations.

A rough estimate for an equivalent noise strength for the cubic GLE would be 
to demand that the reduced noise "temperature" $\sigma^2/V(R_N)$ is in both 
cases the same. This would
require for the cubic GLE at a common $\mu$ of, say, 0.05 that $\sigma$ is by
about a factor of 13 smaller than for the quintic GLE. 

\begin{figure}
\includegraphics[clip=true,width=12cm,angle=0]{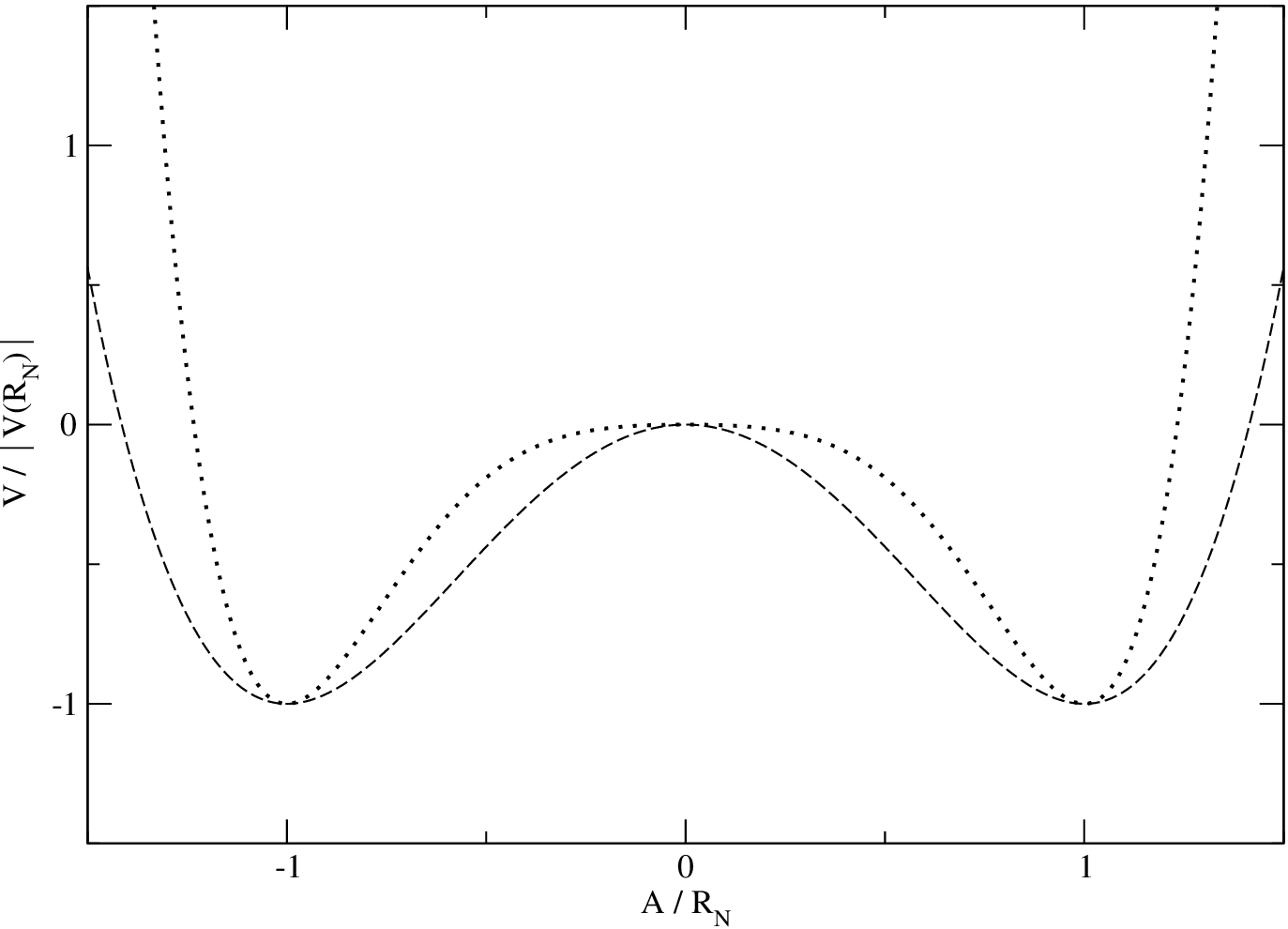}
\caption{\label{figure:V_3-5}
Reduced potentials V (\ref{V_3-5}) corresponding to the real cubic (dashed line)
and quintic (dotted line) GLE. For the cubic case the plot is independent of 
$\mu, g_3$. For the quintic case it depends on the combination 
$\tilde \mu = -g_5 \mu /g_3^2$. Here $\tilde \mu = 0.05$.}
\end{figure}

However, basing the comparison on the requirement that $\sigma^2/ V(R_N)$
is the same for the cubic and quintic case one has to keep in mind that the
curvatures of $V$ around the states $A=0$ and $A=R_N$ which are connected by the
fronts remain different -- cf. Fig.~\ref{figure:V_3-5}. Since these curvatures
around $A=0$ ($A=R_N$) measure the growth (decay) rates of fluctuations around
the respective states it is useful to compare their ratios via a kind of
Ginzburg number $G=|V^{''}(0)|/V^{''}(R_N)$. One has $G_{cubic} = 1/2$
independent of $\mu$ and $g_3$ while 
$G_{quintic} = \tilde \mu\left[ 1 + 4\tilde \mu
+\sqrt{1 + 4\tilde \mu} \right]^{-1}$ with $\tilde \mu = -g_5 \mu /g_3^2$. Thus 
for $\mu=0.05$ and $g_3=1, g_5=-1$ one has 
$G_{cubic} \simeq 23~G_{quintic}$. This largely explains the stronger noise
sensitivity of the cubic GLE for our parameters. In view of it we 
investigated the whole range of $\sigma$ between $10^{-9}$ and $10^{-2}$
for the cubic GLE.

The cubic GLE with additional (but very small) complex coefficients has
previously been investigated, e.g., for noise strengths of about 
$\sigma=1.9\cdot10^{-6}$ in our units of eqs.~(\ref{CGLE}-\ref{f_corr}). The
corresponding noise "temperature" $\sigma^2/V(R_N)$ is about $10^{-8}$ for a 
typical value
of, say, $\mu=0.035$ \cite{BAC}. This noise was found to fit the experimental
results on the noise sustained traveling Taylor vortices under statistically
stationary fronts in the convectively unstable regime
of open Taylor-Couette systems with axial through-flow \cite{BAC}.

%%%%%%%%%%%%%%%%%%%%%%%%%%%%%% Sec IID  %%%%%%%%%%%%%%%%%%%%%%%%
\subsection{Numerical methods}
\label{SEC:num_methods}
%%%%%%%%%%%%%%%%%%%%%%%%%%%%%%%%%%%%%%%%%%%%%%%%%%%%%%%%%%%%%%%%%%
Equation (\ref{CGLE}) was solved numerically with a forward-time, centered-space
method \cite{Pre94} subject to the boundary conditions
\begin{equation}
A(x=0,t)\;=\;0\;=\;A(x=L,t)
\end{equation}
on the complex amplitude. System sizes $L$ were chosen to be sufficiently large
to allow for the establishment of a saturated bulk amplitude. Typically,
a spatial step $dx= 0.4$ was used with a time step of $dt=0.072$. Calculations
were performed for sequences of the paramater $v$ at several values of the control
parameter $\mu$. Most of them 
were done at $\mu=0.05$. The noise source $\eta$ was realized by Gaussian distributed
random numbers of unit variance that were divided by $\sqrt{dt dx}$
to ensure independence of the correlation functions of the discretization.
A test of different pseudo random number generators, namely, L'Ecuyer's method
with Bays-Durham shuffle \cite{Pre94}, \verb/ran3/ \cite{Pre94}, and the \verb/R250/
shift-register random number generator \cite{R250} gave similar results.

After the simulations were started, a sufficiently long time depending on the
parameters, e.g., on the closeness to the convective-absolute threshold had to
be waited until the system relaxed into a statistically stationary state
with time independent averages. Thereafter time averages were evaluated over
several consecutive time intervals and finally averaged.
Within the forward-time integration method $A(x,t)$ remains uncorrelated with
$\eta(x',t)$ at the same time, $<f(\eta)g(A)>\; =\;<f(\eta)><g(A)>$, so that,
e.g., $<A\eta>\;=\;0$ as well as $<q\eta>\;=\;0$. But $<\Omega\eta>\;\ne\;0$. Here
the frequency $\Omega$ (wave number $q$) is defined as a forward-time
(centered-space) difference of the phase (\ref{Omegaq_def}).

%\clearpage
%%%%%%%%%%%%%%%%%%%%%%%%%%%  Sec. III %%%%%%%%%%%%%%%%%%%%%%%%%%%%%%%%
\section{Results}
%%%%%%%%%%%%%%%%%%%%%%%%%%%%%%%%%%%%%%%%%%%%%%%%%%%%%%%%%%%%%%%
The influence of additive noise on the pattern formation process described by
the GLE (\ref{CGLE}) is described in this section.
%%%%%%%%%%%%%%%%%%%%%%%%%%%  Sec. IIIA %%%%%%%%%%%%%%%%%%%%%%%%%%%%%%%%
\subsection{Growth length $\ell$}
\begin{figure}
\includegraphics[clip=true,width=14cm,angle=0]{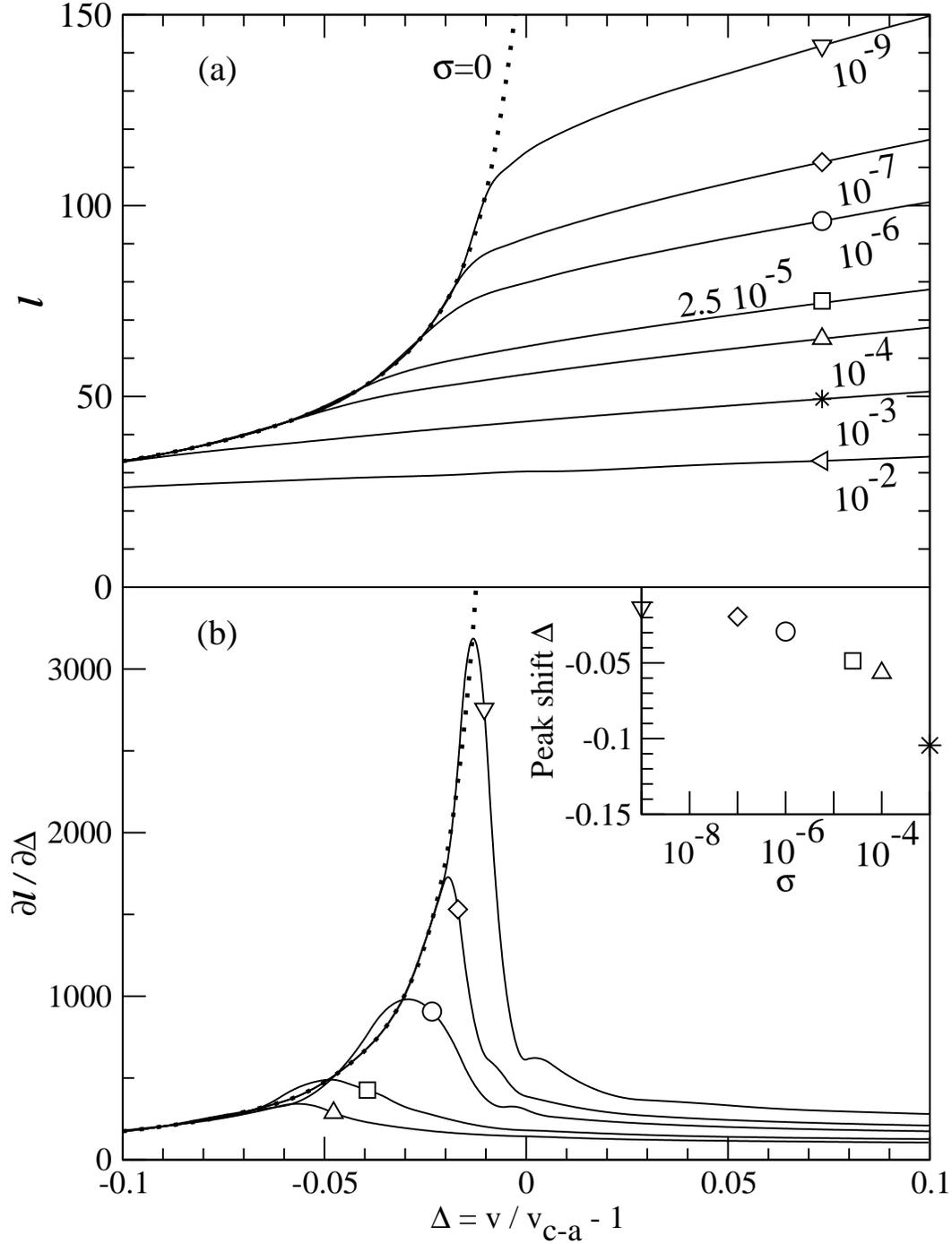}
\caption{\label{figure:l-dl}
Growth length $\ell$ (a) and its derivative $\partial\ell/\partial\Delta$ (b)
versus $\Delta$ near the convective-absolute threshold for the cubic GLE at 
$\mu=0.05$ for
various noise strengths $\sigma$. Inset in (b) shows the variation of the peak of 
$\partial\ell/\partial\Delta$ with $\sigma$.}
\end{figure}

%%%%%%%%%%%%%%%%%%%%%%%%%%%%%%%%%%%%%%%%%%%%%%%%%%%%%%%%%%%%%%%
In Fig.~\ref{figure:l-dl} we show how the growth length $\ell$ of the downstream
pattern occurring in the forced cubic GLE varies with noise strength $\sigma$.
Here $\ell$ is defined by the distance from $x=0$ at which the root-mean square
$\sqrt{<|A|^2>}$ of the fluctuating complex amplitude $A$ reaches half its bulk
value. In the absence of noise $\ell$ diverges at the convective-absolute 
threshold
$v=v_{c-a}$ since there the deterministic pattern is blown out of the system.

For finite $\sigma$ the solution with finite $A$ is noise-sustained in the
convectively unstable regime $\Delta>0$ \cite{D85}. In this regime $\ell$ is
far from the convective-absolute threshold well described by the relation
$\ell\sim - (1 + \sqrt{2\Delta})\ln{\sigma}$ following from a quasilinear
analysis of the cubic GLE \cite{LS97} presented here in an appendix.
However in the vicinity of the threshold
$\Delta\,=\,0$ the growth length $\ell$ obtained from the nonlinear GLE shows a
characteristic crossover to the behavior at $\Delta<0$.

The noise influences also in this absolutely unstable regime, $\Delta<0$,
the finite amplitude solution at least close to threshold:
The curves $\ell(\Delta,\sigma)$ in Fig.~\ref{figure:l-dl}(a) break away from
the
dotted $\ell(\Delta,\sigma=0)$ reference growth length curve at negative
$\Delta$ values
that decrease  with increasing $\sigma$, i.e., further and further away from the
convective-absolute threshold. The associated inflection points can be most 
easily identified
by the maxima in $\partial\ell(\Delta,\sigma)/\partial\Delta$ shown
in Fig.~\ref{figure:l-dl}(b). These peak positions of
$\partial\ell/\partial\Delta$ vary with
$\sigma$ as shown in the inset of Fig.~\ref{figure:l-dl}(b). So the growth
length shows
for the cubic GLE a definite noise sensitivity also in the absolutely unstable
regime.

\begin{figure}
\includegraphics[clip=true,width=13cm,angle=0]{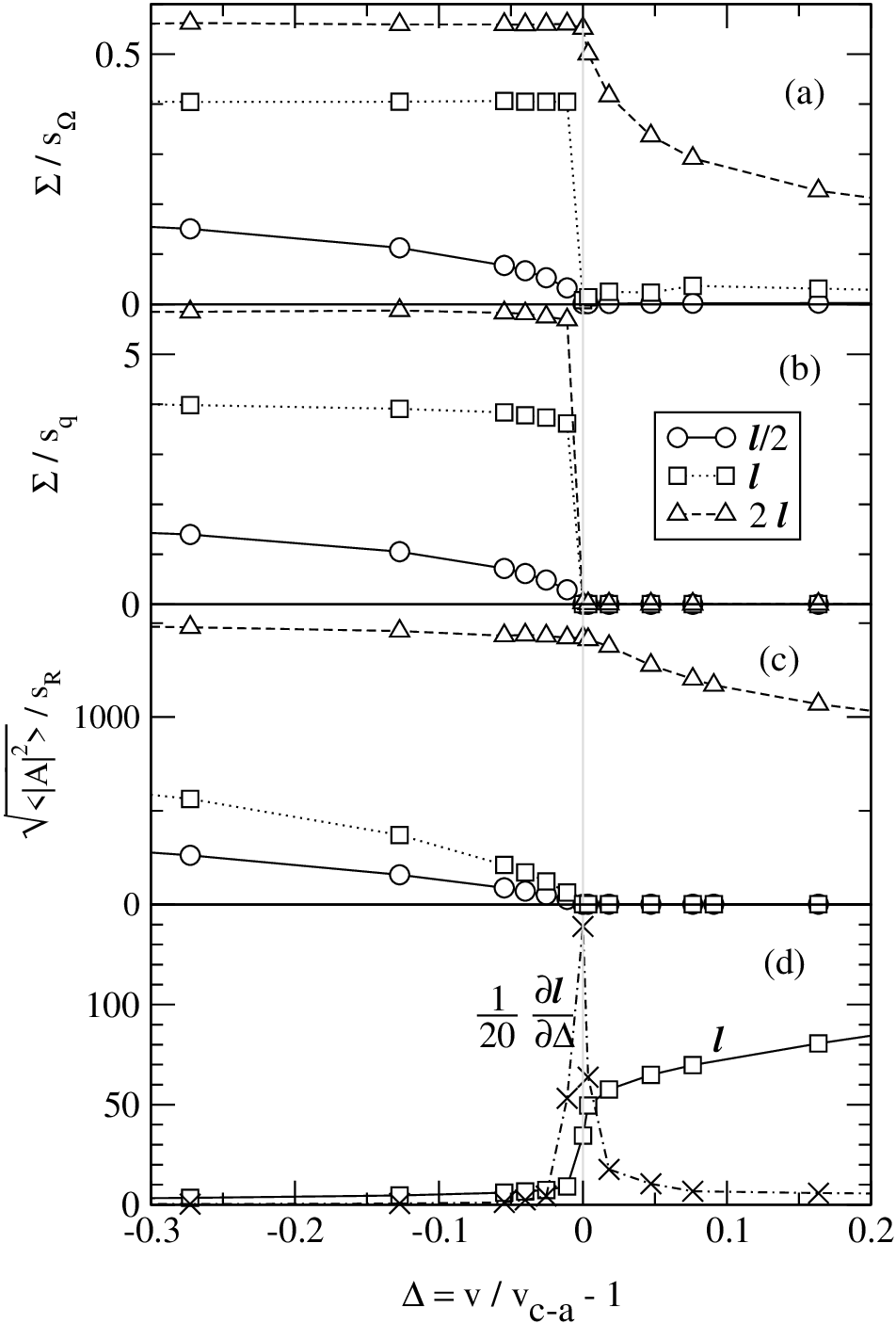}
\caption{\label{figure:l-s_qOm-quin}
Inverse of the standard deviations of frequency $s_\Omega$ (a),
wave number $s_q$ (b), and amplitude modulus $s_R$ (c) for the stochastic quintic 
GLE. Results are
reduced by $\Sigma$ (\ref{Sigma_def}) or $\sqrt{<|A|^2>}$, respectively, and 
plotted as functions of $\Delta$ for
three downstream locations $x=\ell/2$, $\ell$, and $2\ell$. (d) shows the growth
length $\ell$ of $\sqrt{<|A|^2>}$ together with its derivative
$\partial\ell/\partial\Delta$. Piecewise straight lines are guides to the eye.
Parameters are $\mu=0.05$ and $\sigma=10^{-3}$.}
\end{figure}

This sensitivity is significantly smaller in the quintic GLE. This can be seen
by comparing the behavior of the growth length with the fluctuations of
the modulus
$R=|A|$, of the frequency, and of the wave-number (cf, Sec.~\ref{SEC:omega_q}).
To that end we show in Figs.~\ref{figure:l-s_qOm-cub} and ~\ref{figure:l-s_qOm-quin}
$\ell$ and $\partial\ell/\partial\Delta$ together with the inverse of the
standard deviations of the modulus
\begin{equation}
\label{sR_def}
s_R=\sqrt{<R^2> - <R>^2} \, ,
\end{equation}
of the frequency $s_{\Omega}$ (\ref{s_def}), and of the wave-number
$s_q$ (\ref{s_def}) at $\mu$=0.05
as functions of $\Delta$ for the cubic and quintic GLE, respectively.
The noise strengths $\sigma=2.5\cdot 10^{-5}$ and $10^{-3}$, respectively, used for
these figures are roughly equivalent based on the criterion described in
Sec.~\ref{SEC:noise_strength}. However, the potential minima in the cubic case are
broader than in the quintic case -- cf. Fig.~\ref{figure:V_3-5} -- and therefore
the modulus fluctuations in the
former are larger than those in the latter one. This can be seen by comparing the
reduced inverse $\sqrt{<R^2>}/s_R$ in the absolutely unstable regime, $\Delta<0$, of
Figs.~\ref{figure:l-s_qOm-cub}(c) and \ref{figure:l-s_qOm-quin}(c).

The peak position of $\partial\ell/\partial\Delta$ coincides with the drop-off
in the inverse standard deviations $1/s$. For the cubic GLE
(Fig.~\ref{figure:l-s_qOm-cub}) it occurs
at $\Delta$=-0.049, thus being shifted significantly into the
absolutely unstable regime while that of the quintic GLE 
(Fig.~\ref{figure:l-s_qOm-quin}) remains at $\Delta$=0.

As an aside we mention that for the quintic GLE at a subcritical growth
parameter of, say, $\mu$=-0.05 the behavior of the growth length $\ell$ and
of $\partial\ell/\partial\Delta$ is for $\Delta<0$ similar to the one shown in
Fig.~\ref{figure:l-s_qOm-quin}(d) for $\mu$=0.05. For $\mu <0, \Delta > 0$ we
did not find a noise sustained large-amplitude solution.

%%%%%%%%%%%%%%%%%%%%%%%%%%%  Sec. IIIB %%%%%%%%%%%%%%%%%%%%%%%%%%%%%%%%
\subsection{Frequency and wave-number correlations}
\label{SEC:omega_q}
\begin{figure}
\includegraphics[clip=true,width=13cm,angle=0]{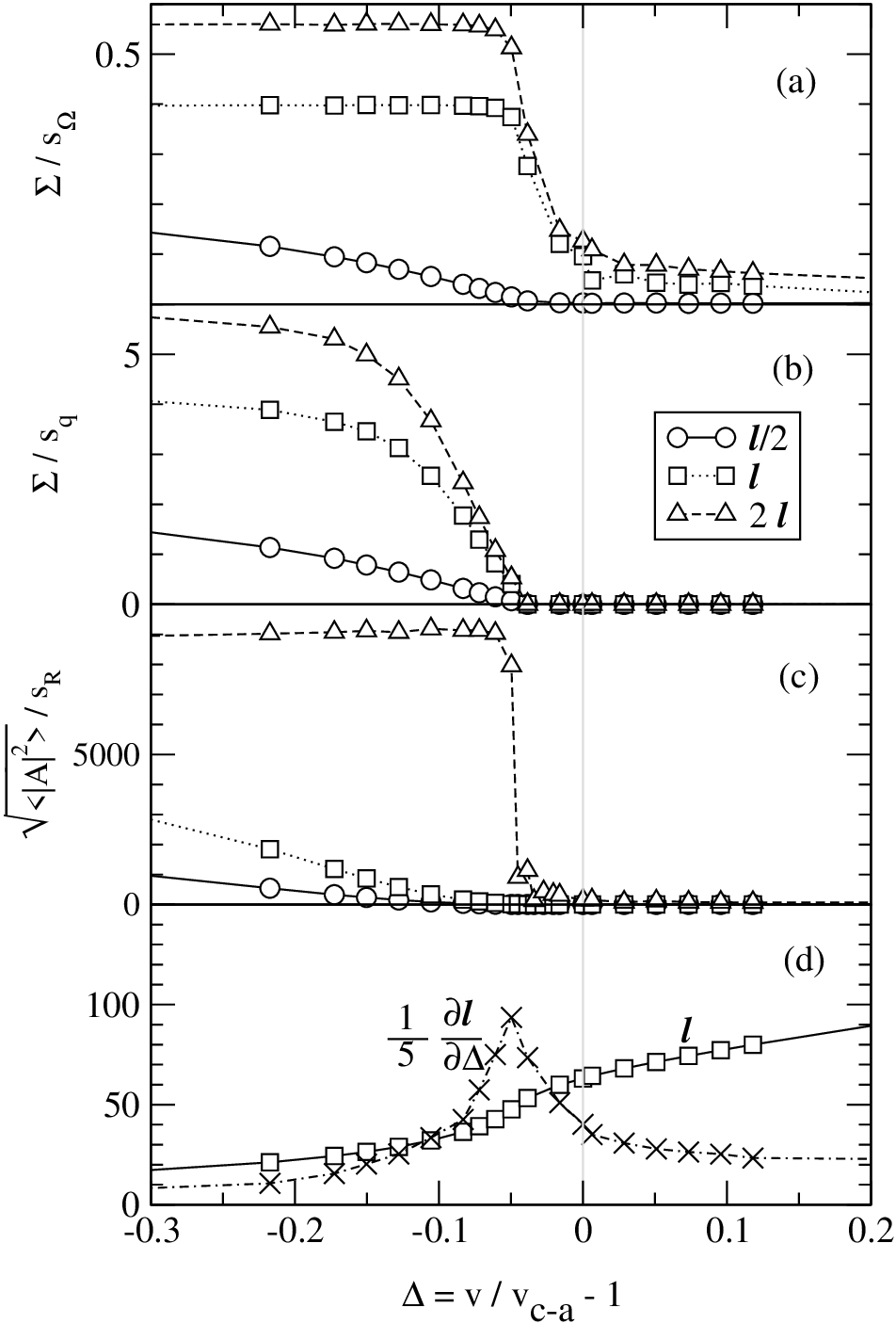}
\caption{\label{figure:l-s_qOm-cub}
Inverse of the standard deviations of frequency $s_\Omega$ (a), 
wave number $s_q$ (b), and amplitude modulus $s_R$ (c) for the stochastic cubic
GLE. Results are
reduced by $\Sigma$ (\ref{Sigma_def}) or $\sqrt{<|A|^2>}$, respectively, and
plotted as functions of $\Delta$ for
three downstream locations $x=\ell/2$, $\ell$, and $2\ell$. (d) shows the growth
length $\ell$ of $\sqrt{<|A|^2>}$ together with its derivative
$\partial\ell/\partial\Delta$. Piecewise straight lines are guides to the eye.
Parameters are $\mu=0.05$ and $\sigma=2.5\cdot 10^{-5}$.}
\end{figure}

%%%%%%%%%%%%%%%%%%%%%%%%%%%%%%%%%%%%%%%%%%%%%%%%%%%%%%%%%%%%%%%
 Previous investigations of the forced cubic GLE in the bulk
part of the solution at far downstream locations
$x\gg\ell$ showed for different but small noise strengths that frequency 
fluctuations are in the absolutely unstable regime much smaller than in the
convectively unstable regime \cite{BAC}. In order to study this question of
the noise sensitivity in both regimes
we have investigated in more detail the frequency and wave-number
fluctuations at
$x=\ell/2,\ell$, and $2\ell$. The results are shown in Fig.~\ref{figure:l-s_qOm-cub}
for the cubic GLE and in Fig.~\ref{figure:l-s_qOm-quin} for the quintic GLE.
Before
we discuss them we first present some basic properties of the phase fluctuations
as described by the forced GLE (\ref{CGLE}).

The phase fluctuations $\Phi$ of the complex amplitude (\ref{A_ReIm}) define
the frequency $\Omega$ and the wave number $q$
\begin{equation}
\Omega\;=\;\dot \Phi\;=\;  \Im\left(\frac{\dot A}{A}\right)\;,  \qquad     %%{\cal I}m \frac{\dot A}{A}\;,  \qquad
q\;=\;\Phi'\;=\;\Im\left( \frac{A'}{A}\right)\;,
\label{Omegaq_def} \end{equation} respectively. Here dot (prime) denotes temporal
(spatial) derivative. The growth rate $\kappa$ of the modulus is given by
\begin{equation} \kappa\;=\;\frac{R'}{R}\;=\;\Re\left( \frac{A'}{A}\right)\; .
\end{equation}
By means of Eq.~(\ref{CGLE}) the frequency can be expressed as
\begin{equation}
\Omega\;=\; (2\kappa -v) q + q' + \frac{\sigma}{R^2}  \Im\left(\eta
A^*\right)\;.
\label{Omega_final}
\end{equation}
This relation holds for the cubic as well as for the quintic GLE with real
coefficients. By squaring and averaging Eq.~(\ref{Omega_final}) one
gets the correlation functions
\begin{eqnarray}
<\Omega^2> &+& v^2 <q^2> + 2v <\Omega q> +
<q^{\prime\; 2}> - 2 <\Omega q'> - 2 v <qq'>
\nonumber \\
&-& 4 v<\kappa q^2> + 4 <\kappa^2 q^2>
- 4 <\kappa \Omega q> + 4 <\kappa q q'>\nonumber \\\;&\simeq&\;
\frac{\sigma^2<|\eta|^2>}{2<R^2>} \; .
\label{Omega2_full}
\end{eqnarray}
On the r.h.s. we have used the fact that within our forward-time integration
method $A$ remains uncorrelated with $\eta$ at the same time and we have
approximated  $<1/R^2>$ by $1/<R^2>$.

Given that $<|\eta(t,x)|^2>=2 /dx dt$ in our finite difference
simulation it is convenient to scale all correlations in Eq.~(\ref{Omega2_full})
by the quantity
\begin{eqnarray}
\Sigma^2\;=\;\frac{\sigma^2}{R_N^2}\,\frac{1}{dx dt} \; , \qquad
R_N^2=\left\{ \begin{array}{ll} \mu &\mbox{cubic GLE} \\
\frac{1}{2} + \sqrt{\mu + \frac{1}{4}}\quad & \mbox{quintic GLE}
\end{array}\right.
\label{Sigma_def}
\end{eqnarray}
thereby removing the singularities from the reduced correlation functions. For
example one finds that
\begin{equation}
\frac{<(\Omega + v q - q')^2>}{\Sigma^2} \simeq \frac{R_N^2}{<R^2>}\;.
\label{Om2vsSigma}
\end{equation}
Here we have neglected the second line in Eq.~(\ref{Omega2_full}) since all
correlations in Eq.~(\ref{Omega2_full}) involving the growth rate $\kappa$ are
very small.

$<\Omega^2>$ is typically two orders of magnitude larger than $<q^2>$ in the
absolutely unstable regime, $\Delta < 0$, -- cf.
Figs.~\ref{figure:l-s_qOm-cub} and \ref{figure:l-s_qOm-quin} discussed further
below. There the only contributions to
Eqs.~(\ref{Omega2_full},\ref{Om2vsSigma}) of the same order as $<\Omega^2>$ are
$<\Omega q'>$ and $<q'^2>$ -- all the other correlations can be neglected --
and furthermore $<\Omega q'> \simeq <q'^2>$. Thus,
\begin{equation}
<\Omega^2>\;\simeq\;\Sigma^2 + <q'^2>
\end{equation}
in the bulk part of the system with saturated amplitude where
$<R^2> \simeq R_N^2$.
However, in the convectively unstable regime, $\Delta > 0$, with much larger
phase fluctuations the situation is more complex. Here $<q^2>$ is larger than
$<\Omega^2>$ except for the upstream region where the reverse holds.

In Fig.~\ref{figure:l-s_qOm-cub} and Fig.~\ref{figure:l-s_qOm-quin} we show
the inverse of the standard deviations
\begin{eqnarray} \label{s_def}
s_\Omega\;&=&\;\sqrt{<\Omega^2> - <\Omega>^2}\;, \qquad
s_q\;=\;\sqrt{<q^2> - <q>^2}\;,
\end{eqnarray}
reduced by $\Sigma$ (\ref{Sigma_def}) for the cubic and quintic GLE,
respectively, as functions of $\Delta$ for $x=\ell/2,\ell$, and $2\ell$.
For the parameters shown in Fig.~\ref{figure:l-s_qOm-cub} and
Fig.~\ref{figure:l-s_qOm-quin} the mean frequency $<\Omega>$ as well as the
mean wave number $<q>$ are negligible.
Plotting the inverse of $s_\Omega$, $s_q$, and $s_R$  allows to visualize the small
fluctuations in the absolutely unstable regime better than in a direct plot
of, say, $s_\Omega^2$. Such plots for $s_\Omega^2$ have been presented
previously for the small noise strengths occurring in Taylor-Couette
experiments \cite{BAC}. On the lower level of resolution inherent in this data
presentation these results show similar behavior as 
ours. However, plotting $1/s_\Omega$ instead allows 
to identify more clearly the crossover behavior
from the parameter regime with small fluctuations to the one with large ones.

The $\Delta$-variations of $1/s_\Omega$, $1/s_q$, $1/s_R$, and of
$\partial\ell/\partial\Delta$ indicate that this transition is
shifted to negative $\Delta$, i.e.~into the absolutely unstable regime.
A similar result for the transition between deterministic and noise
sustained standing wave solutions of complex coupled cubic GLE's was deduced from
the behavior of the second moments of the frequency and wave-number power spectra
of the fluctuating amplitudes \cite{NWS96}: With decreasing $\mu$ the 
correlation length defined via the time average of the second moment of the 
Fourier spectrum of $A(k,t)$ begins to decrease towards values
characteristic for noise-sustained structures in the convectively unstable
regime clearly before $\mu_{c-a}$ is reached when noise is present. Similarly
the width of the frequency power spectrum starts to increase with decreasing
$\mu$ already above the convective-absolute threshold $\mu_{c-a}$ \cite{NWS96}.

 However, the variation of $1/s_\Omega$ with $\Delta$ shows for the cubic case in 
Fig.~\ref{figure:l-s_qOm-cub} a broader crossover
interval between large frequency fluctuations in the convectively unstable
regime at $\Delta >0$ and small frequency fluctuations in the absolutely
unstable regime at $\Delta <0$ than the curves $1/s_q$ and $1/s_R$ for wave-number
and modulus fluctuations. The $\Delta$-value at which $1/s_q$ and  $1/s_R$ drop 
down towards zero
agrees quite well with the peak location of $\partial\ell/\partial\Delta$. The
latter moves with increasing noise strength further into the absolutely unstable
regime as shown, e.g., for the cubic GLE in the inset of
Fig.~\ref{figure:l-dl}(b).

The variations of $s(\Delta)$ with $\Delta$ at different
downstream locations $x=\ell/2,\ell$, and $2\ell$ are similar to each other:
with $\Delta$ becoming more negative, i.e., further and further into
the absolutely unstable regime the fluctuations $s_\Omega$ and $s_q$ become
constant at levels that depend on the measuring location -- the closer to the
inlet where $R$
becomes smaller the larger are the fluctuations. This behavior is reflecting the
relation $s_\Omega \sim s_q \propto R^{-1}$  that can be read off directly from
Eq.~(\ref{Omega2_full}).

\begin{figure}
\includegraphics[clip=true,width=13cm,angle=0]{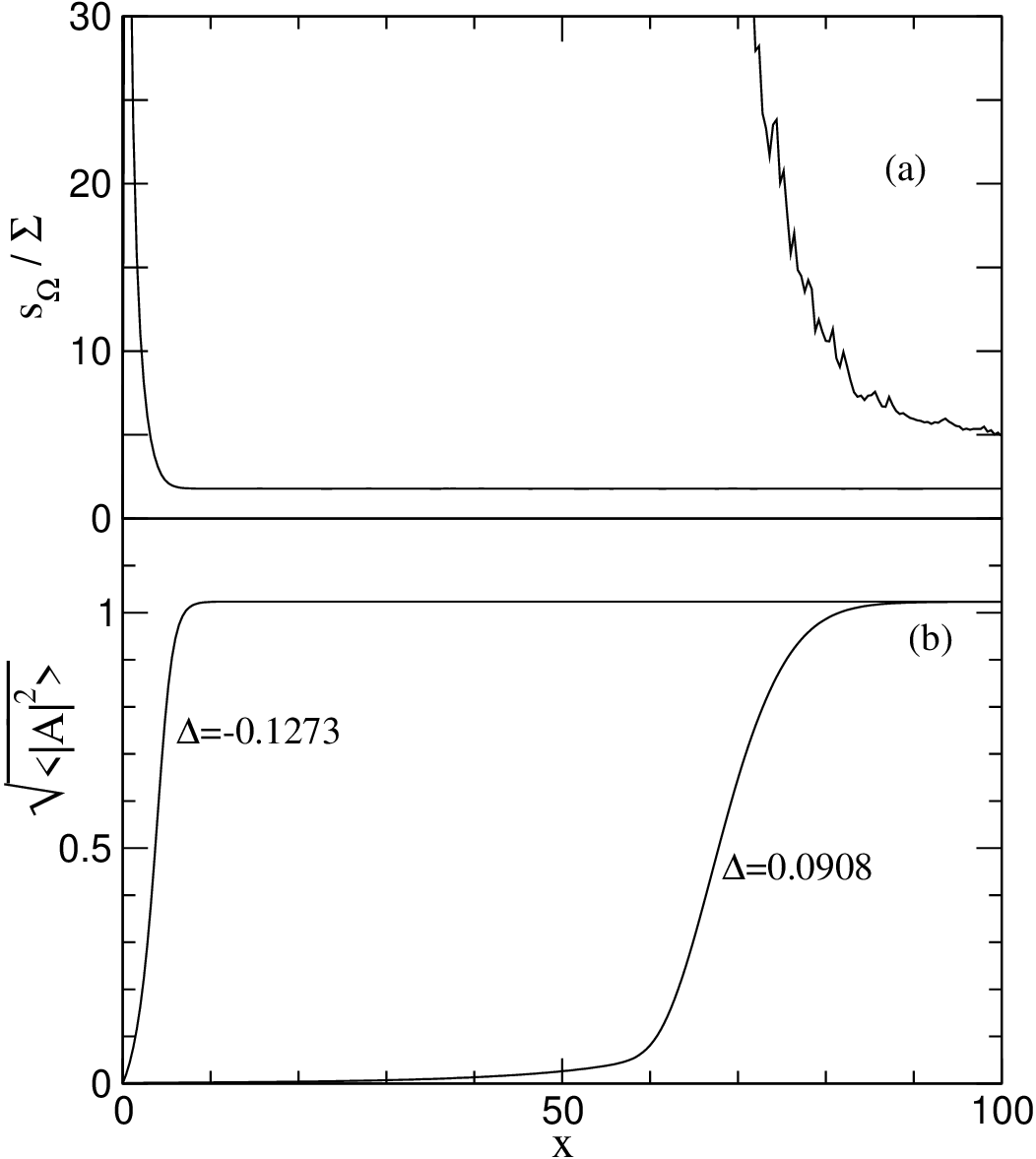}
\caption{\label{R-s_om-x}
Spatial variation of the standard deviation $s_\Omega$ of the frequency reduced by
$\Sigma$ (a) and of $\sqrt{<|A|^2>}$ (b) for the quintic GLE in the absolutely and
convectively regime at $\Delta = -0.1273$ and $\Delta = 0.0908$, respectively.
After the integration time of $T=5\cdot10^5$ used in this plot $s_\Omega$ was for 
$\Delta = 0.0908$ not yet fully stationary in the growth region of
$\sqrt{<|A|^2>}$. Parameters are $\mu=0.05$ and $\sigma=10^{-3}$.} 
\end{figure}

The downstream reduction of the variance $s_\Omega$ of the frequency fluctuations
with increasing distance from the inlet and with increasing amplitude along the
front is shown in Fig.~\ref{R-s_om-x} for the quintic GLE. There we compare the
behavior
of $s_\Omega$ together with the front profiles of $\sqrt{\langle |A|^2 \rangle}$
in the absolutely and in the
convectively unstable regime close to the threshold $\Delta\,=\,0$ for $\mu=0.05$.

%\clearpage
%%%%%%%%%%%%%%%%%%%%%%%%%%%%%%%%%%%%%%%%%%%%%%%%%%%%%%%%%%%%%%%%%%%%%%%%%%%%%%%%%
\section{Conclusion}
%%%%%%%%%%%%%%%%%%%%%%%%%%%%%%%%%%%%%%%%%%%%%%%%%%%%%%%%%%%%%%%%%%%%%%%%%%%%%%%%%
We have studied numerically the influence of small additive noise on pattern formation
near a forwards and near an inverted bifurcation as described by a cubic and
quintic GLE, respectively, when a finite group velocity $v$ can blow the
finite-amplitude part out of the system, i.e., in the vicinity of the
so-called convective-absolute instability at $\Delta=v/v_{c-a}(\mu)-1=0$. The 
front that connects the inlet condition
$A(x=0)=0$ to the finite-amplitude downstream bulk part 
$\langle |A|^2 \rangle \simeq R_N^2$ is for the cubic GLE
more sensitive to the applied noise strength than for the quintic case.
This is partly related to the different magnitudes of the curvatures of the 
deterministic GLE potentials around the states $A=0$ and $A=R_N$: the resulting 
growth enhancement
of fluctuations near $A=0$ is larger in the cubic than in the quintic case and
in addition the damping of fluctuations near $A=R_N$ is smaller in the cubic than 
in the quintic case. 

In the cubic case the transition between the regimes of small and large
fluctuations of amplitude, frequency, and wave number is shifted to a negative
$\Delta$ into the absolutely unstable regime. Simultaneously the pattern growth
length $\ell(\Delta)$
has there a characteristic inflection point that shows up as a peak in
$\partial\ell/\partial\Delta$. In the quintic case all this occurs at the
unshifted convective-absolute threshold $\Delta=0$. Common to both cases is that the
fluctuations decrease along the front in both regimes with growing pattern
amplitude $\sqrt{<|A|^2>}$.

For negative subcritical amplitude growth rates, $\mu < 0$, we did not find
noise-sustained,
large-amplitude, backwards bifurcating patterns when $\Delta$ is positive: the
nonlinear deterministic front dynamics of the quintic GLE blows any
large-amplitude part downstream away from the inlet where $A=0$
and eventually any finite system is filled only with small-amplitude fluctuations
of $A$ around the stable fixed point $A=0$ of the unforced system.

\acknowledgments
Discussions with B. Neubert and his contributions to an early
stage of this research project are gratefully acknowledged.
One of us (A.~S.) acknowledges the hospitality of the Universit\"at
des Saarlandes.

%%%%%%%%%%%%%%%%%%%%%%%%%%%%%%%%%%%%%%%%%%%%%%%%%%%%%%%%%%%%%%%%%%%%%%%%%%%%%%%%%
\begin{appendix}*
\section{}  %{Noise dependence of the growth length}
%%%%%%%%%%%%%%%%%%%%%%%%%%%%%%%%%%%%%%%%%%%%%%%%%%%%%%%%%%%%%%%%%%%%%%%%%%%%%%%%%
Here we estimate the noise dependence of the downstream growth length $\ell$ 
of the nonlinear structure in the convectively 
unstable regime of the cubic GLE where this structure is noise sustained. 
To that end we approximate 
$\ell$ by the length where the mean squared amplitude 
$C_{lin}(x) = \langle |A_{lin}(x)|^2\rangle$ of the {\em linear} GLE has grown 
from the inlet value $A(x=0)=0$ to, say, one half of the nonlinearly saturated 
bulk value $\langle |A|^2\rangle \simeq \mu/2$. So we solve the equation
\begin{equation} \label{lin-cub}
C_{lin}(x=\ell)\;=\;\frac{1}{2} \mu\
\end{equation}
for $\ell$. Actually the linear solution may not hold there anymore. But as it 
will become obvious 
below the result is roughly independent of the coefficient chosen 
in Eq.~(\ref{lin-cub}) so also  smaller numbers than $\frac{1}{2}$ could be
chosen here for a characteristic growth length. 
 
We evaluate the equal-time correlation $C_{lin}(x)$ via the frequency integral
of the spectrum $C_{lin}(x,\omega)$ of the time-displaced autocorrelation 
function of fluctuations of $A_{lin}$ at the same downstream position $x$. 
For large downstream distances $x$ from the inlet this spectrum is given 
by \cite{LS97}
\begin{equation}  \label{C_asympt}
C_{lin}(x,\omega) \;=\; \frac{-\sigma^2}{2 | K_1^* - K_2 |^2 }   
\left(\frac{1}{\Im K_1} + \frac{1}{\Im K_2} \right) e^{-2 \Im K_1 x}
\end{equation}
with 
\begin{equation} \label{eigenvalues}
K_{\binom{1}{2}} \;=\; \pm i \sqrt{\mu_{c-a} - \mu -i\omega} -i \sqrt{\mu_{c-a}}\;. 
\end{equation}
This spectrum (\ref{C_asympt}) is strongly peaked at the center, $\omega=0$, of 
the band of  
modes, $-2\sqrt{\mu\mu_{c-a}} < \omega < 2\sqrt{\mu\mu_{c-a}}$, that are 
amplified in the convectively unstable regime. Thus, the aforementioned frequency
integral may be approximated by
\begin{eqnarray}
C_{lin}(x) = \int_{-\infty}^\infty\frac{d\omega}{2\pi} C_{lin}(x,\omega)
\sim \sqrt{\mu\mu_{c-a}} C_{lin}(x,\omega=0) = 
\frac{\sigma^2}{4\sqrt{\mu}}\exp^{2 i K_1\left(\omega=0\right)x}\;.
\end{eqnarray} 
The last equality follows from Eq.~(\ref{C_asympt}) at $\omega=0$.
Applying now the condition (\ref{lin-cub}) one obtains
\begin{equation}
\ell\;\sim\;\frac{1}{iK_1(\omega=0)} \ln{\frac{2^{1/2}\mu^{3/4}}{\sigma}}\;.
\end{equation}
Using $\mu_{c-a}/ \mu = (1+\Delta)^2$ in Eq.~(\ref{eigenvalues}) one sees that 
$i K_1(\omega=0)=\sqrt{\mu}\left[1 - \sqrt{2\Delta} + {\cal O}(\Delta)\right]$
for  $\Delta \ll 1$ so that finally at fixed $\mu$  
\begin{equation}
\ell\;\sim\;-\left[1 + \sqrt{2\Delta} + {\cal O}(\Delta)\right]
\left(\ln{\sigma} + const \right)\; .
\end{equation}

\end{appendix}

\clearpage
%%%%%%%%%%%%%%%%%%%%%%%%%%%% Fig. 1  %%%%%%%%%%%%%%%%%%%%%%%%%%%%
%%%%%%%%%%%%%%%%%%%%%%%%%%%% Fig. 2  %%%%%%%%%%%%%%%%%%%%%%%%%%%%
%%%%%%%%%%%%%%%%%%%%%%%%%%%% >>>New<<< Fig. 3  %%%%%%%%%%%%%%%%%%%%%%%%%%%%
 %%%%%%%%%%%%%%%%%%%%%%%%%%%% Fig. 4  %%%%%%%%%%%%%%%%%%%%%%%%%%%%

%%%%%%%%%%%%%%%%%%%%%%%%%%%% Fig. 5 %%%%%%%%%%%%%%%%%%%%%%%%%%%%
%%%%%%%%%%%%%%%%%%%%%%%%%%%% Fig. 6  %%%%%%%%%%%%%%%%%%%%%%%%%%%%
%%%%%%%%%%%%%%%%%%%%%%%%%%%% Fig. 7  %%%%%%%%%%%%%%%%%%%%%%%%%%%%
%%%%%%%%%%%%%%%%%%%%%%%%%%%%%%%%%%%%%%%%%%%%%%%%%%%%%%%%%%%%%%%%%%

\end{document}